\documentclass{article}

\usepackage[]{graphicx}
\usepackage{amsmath}

\newcommand{\Sh}{Schr\"odinger{ }}
\newcommand{\etal}{et al. }
\newcommand{\mca}[1]{\mathcal{#1}}
\newcommand{\bscal}[1]{\bs{\mathcal{#1}}}
\newcommand{\bs}[1]{\boldsymbol{#1}}

\newcommand{\braket}[3]{\left\langle #1 \left\vert #2
            \right\vert #3 \right\rangle}

\newcommand{\ket}[1]{\left\vert #1 \right\rangle}
\newcommand{\fuve}[1]{\left(\bs{#1}\right)}
\newcommand{\fues}[1]{\left(#1\right)}

\newcommand{\fusa}[3]{_{\alpha,\bs{#1}}\fues{#2,\bs{#3}}}

\newcommand{\yav}[1]{\left[#1\right]}

\newcommand{\Eq}[1]{eq. (\ref{#1})}

\newcommand{\pr}{^{\prime}}
\newcommand{\nas}[1]{\nabla_{\bs{#1}}}
\newcommand{\sa}{_{\alpha}}

\title{Hall conductivity as a topological invariant}
\author{A. Kunold$^1$, M. Torres$^2$\\
$^1$Departamento de Ciencias B\'asicas,\\
Universidad Aut\'onoma Metropolitana Azcapotzalco,\\
Av. San Pablo 180, Col. Reynosa Tamaulipas,\\
Azcapotzalco, M\'exico D.F. 02200, M\'exico,\\
e-mail: {\sf akb@correo.azc.uam.mx}, Phone: +53\,18\,93\,81\\
$^2$Instituto de F\'{\i}sica,\\
Universidad Nacional Aut\'onoma de M\'exico,\\
Apdo. Postal 20-364, M\'exico D.F. 01000, M\'exico,\\
e-mail: {\sf torres@fisica.unam.mx.}, Phone: 56\,22\,51\,44}

\begin{document}

\maketitle

\begin{abstract}
The object of the present work is to study the quantum Hall effect
through its symmetries and topological aspects.
We consider the model of an electron moving in a two-dimensional
lattice in the presence of applied in-plain electric field and
perpendicular magnetic field.
We refer to this as the two dimensional electric-magnetic
Bloch problem (EMB). The Hall conductivity  quatizations beyond
the linear response approximation is analyzed. 
\end{abstract}

\section{Introduction}
The  quantization of the Hall conductance ($\sigma_H$) in integer (IQHE)
and fractional values \cite{Klit1} is an astonishing and unexpected result.
For the IQHE,  the exact quantization of the  Hall conductivity was
explained based on a topological argument by Thouless
et al. \cite{Thou1}; $\sigma_H$ in units of $e^2/h$ is a
topological invariant, the so-called first Chern number \cite{Avron1},
which can only take integer values 
if the Fermi level lies in a gap between Landau levels.
The precense of a  periodic potential broadens each Landau level in
a series of minibands, separated by the corresponding minigaps.
Contrary  to the obvious suggestion that each  subband carries a
fraction of $e^2/h$,  Thouless et al. demonstrated that the contribution
to $\sigma_H$ of each filled miniband  is also an integer multiple of $e^2/h$.
Clear experimental evidence for the internal structure of the
Hofstadter butterfly spectrum has only been found very recently
in the measurement of  $\sigma_H$ for lateral superlattices \cite{Klit2}. 
The work of Thouless et al. makes use of the Kubo linear response theory.
It is of our interest to consider the structure of the Hall current
when a finite electric field ($\boldsymbol{E}$) is applied, in particular we
shall be able to demonstrate that even in this case the Hall conductivity
remains an integer multiple of $e^2/h$.  From the study of electron  wave packet
propagation  in similar physical conditions;  it has been observed the
existence of  transitions from extended  to  localized   states  
when  the electric field  switches 
from commensurate to incommensurate  orientations relative  to the lattice \cite{kun2}.
One interesting feature of our
model is that it can be utilized in order to explore the conductivity behavior
as a function of the electric field orientation \cite{CondMat1}.
\section{Bloch-Floquet Wave Functions}
We consider the EMB problem given by the time dependent \Sh equation
\begin{equation}\label{shr1}
S\ket{\alpha, \boldsymbol{k}}=
\yav{\frac{1}{2 m^*}\fues{\Pi_x^2+\Pi_y^2}+V-\Pi_0}
\ket{\alpha, \boldsymbol{k}}=0.
\end{equation}
Here  $m^*$ is the effective  electron mass,  $ \Pi_{\mu}=p_{\mu}+A_{\mu}$,
$p_{\mu}= (\imath \hbar \partial/\partial t, - \imath \hbar \nabla )$,
the vector and scalar potentials, $A_\mu=(\phi, {\bs A})$,  are selected to
yield: $\nabla \times \mathbf{A}=B \hat{k}$,  and
$\mathbf{E}=-\nabla \phi -\partial \mathbf{A}/\partial t$.  For simplicity
we shall consider a periodic square lattice:
$V\left(x,y\right)=V_0\left[\cos\left(2\pi  x/a \right) +
 \cos\left(2\pi y/a\right)\right].$ The energy and  lengths are  measured in units of 
$ \hbar\omega_c=\hbar e B/m^*,$ and $ \ell_0=\sqrt{\hbar/e B}$,
respectively, where  $\omega_c$ is the cyclotron frequency and $ \ell_0$
the magnetic length,  which is equivalent to setting
$\hbar=e=m^*=1$.

A  symmetry group of $S$ is given by the electric-magnetic
translation group
\cite{Zak1},  which consists of
a combination of space translations and time evolution operations compensated by  a gauge
transformation. In general, the two magnetic translations and the electric-evolution operators do not
commute amongst
themselves, unless three commensurate conditions  are imposed\cite{CondMat1}:
(1) The direction of the electric field, given by the angle $\theta$ with respect to the lattice axis,
is commensurate, i.e. $\theta=\arctan m_1/m_2$ where $m_1$ and $m_2$
are relative prime numbers. Hence an extended lattice along the electric field preserves a spatial periodictity, the  extended lattice parameter is  $b=a\sqrt{m_1^2+m_2^2}$. (2) The magnetic flux through each unit cell of
the extended lattice per flux quanta is a rational number $p/q=1/\sigma$
 (3) Time in the evolution operator is restricted to multiple values of $\tau=b/v_D$,  where
$v_D=E/B$ is the drift velocity of the electron.

The eigenstate base of this group is given by the Bloch-Floquet wave
functions(BF) \cite{CondMat1}
\begin{eqnarray}\label{flo4}
 \varphi_{\alpha, \bs{k}}  \left( t, \bs{x} \right)&=&
e^{\imath \bs{k}\cdot \bs{x}-\imath \mca{E}t} \, \, 
u_{\alpha,\bs{k}}  \left( t, \bs{x} \right) \, \\
u_{\alpha,\bs{k}} \left( t, \bs{x} \right)  &=&
\frac{1}{\sqrt{2\pi}}  e^{-\imath x  (- y/2 + k_x )  }
e^{\imath E t x/2   }
\sum_{\mu l m} i^{\mu}  b_{m}^{\mu}
e^{- \imath qbElt }
 e^{\imath \sigma b\fues{k_1 - y}\fues{m+pl}} 
 \phi_{\mu} , 
\end{eqnarray}
here  $\phi_{\mu}\equiv\phi_{\mu}\fues{x- k_2  + 2\pi\frac{m+pl}{b}}$
is the harmonic oscillator wave function and  $b_{m}^{\mu}$ is a coeficient that satisfies a finite difference equation. For the particular limit in which $\bs E =0$ and   the Landau inter-level couplings are  neglected, the equation coincides with the Harper equations; results for the generalized Harper equation  are reported in reference \cite{CondMat1}.

Besides  the finite difference equation,  
a useful alternative form of the generalized Harper's equation 
in the presence of electric field can be derived within the present formalism, it reads 
\begin{equation}\label{pro3}
\fues{\mca{E}-\imath \bs{E}\cdot\nas{k}}\ket{\alpha, \boldsymbol{k}}
=H\ket{\alpha, \boldsymbol{k}} \, ,
\end{equation}
where $H$ is the Harper's or generalized Harper's  hamiltonian  \cite{CondMat1}.
\section{Hall conductivity}\label{hall}
Using the above described states, the current density operator takes the form 
\begin{equation}\label{cor0}
\bs{\hat{J}}= 
\sum_{\alpha,\alpha\pr}\int \frac{d^2k\pr }{ (2\pi)^2} \int {d^2k \over  (2\pi)^2} \;
b^\dag\sa\fuve{k\pr}b\sa\fuve{k}
\braket{\alpha\pr,\bs{k}\pr}{\bs{\Pi}}{\alpha,\bs{k}},
\end{equation}
where $b^\dag\sa\fuve{k}$ and $b\sa\fuve{k}$ are
the usual Fermi creation and  annihilation operators. 
At low temperatures, it can be proved  that  the electric  current  becomes
\begin{equation}\label{tom1}
\bs{J}=
\left\langle \Psi\right\vert \bs{\hat{J}}\left\vert\Psi\right\rangle
=  \sum_{\alpha\le \alpha_F}
\int {d^2k \over  (2\pi)^2} \;
\yav{\nas{k}\mca{E}\sa\fuve{k}
-\bs{E}\cdot\nas{k}\bscal{A}\sa\fuve{k}},
\end{equation}
where $\bscal{A}^{\alpha}\fuve{k}=  (2 \pi)^2/(q a^2)
\left\langle u\fusa{k}{t}{x}\right\vert
\imath \nas{k}\left\vert u\fusa{k}{t}{x}\right\rangle$. Altough it is clear that  $\bscal{A}^{\alpha}\fuve{k}$ can be  identified with the 
 Berry connection, we  notice that there is not any adiabatic approximation involved, the connection  $\bscal{A}^{\alpha}\fuve{k}$
  is evaluated using the states   that exactly solve \Eq{pro3}.
The first term on the right hand side of  the previous equation
represents the usual group velocity contribution. Additionally there is a novel contribution arising
from  the gradient of the Berry connection along the longitudinal electric
field direction.
Considering the diagonal matrix elements obtained from   eq. (\ref{pro3}), it yields
$\mca{E}\sa\fuve{k}=\Delta\sa\fuve{k}
+\bs{E}\cdot\bscal{A}\sa\fuve{k}-Ek_y$
with
$\Delta\sa\fuve{k}=\langle u_{\alpha,\bs{k}}\vert H\vert u_{\alpha,\bs{k}}\rangle$.
Substituting this results into eq. (\ref{tom1}),   and considering completly filled subbands, it can be readily demonstrated that the current takes the form 
\begin{equation}\label{xyz1}
\bs{J}= 
{\nu_F \over 2 \pi p} E\bs{e}_T-\frac{1}{2 \pi p }\bs{E}\times 
\sum_{\alpha\le \nu_F} \int \frac{d^2k}{2\pi}\;
\bs{\Omega}\sa\fuve{k} \, , 
\end{equation}
here $\bs{e}_T$ is a unit vector transverse to $\bs E$ and  the Berry curvature is given by
$\Omega \sa\fuve{k}=\nas{k}\times\bscal{A}\sa\fuve{k}$.
Equation  (\ref{xyz1}) shows that the longitudinal magnetoresistance 
exactly cancels, regardless of the value of $\bs E$. On the other hand, 
the Hall conductance  can be read  from  \Eq{xyz1}  and  written (restoring units)
in a compact form as 
\begin{equation}\label{cnd4}
\sigma_H=    \frac{e^2}{h} \sum_{\alpha\le\nu_F}  \sigma\sa \, , \,\,\,
\sigma\sa   = \frac{1}{ p} \left( 1 - q \, \eta\sa  \right) \, ,
\end{equation}
where $\eta\sa$ is the $\alpha$ subband contribution to the conductance given as
\begin{equation}\label{www5}
\eta\sa = {1 \over q} \int\frac{d^2k}{2\pi}
\yav{\Omega\sa\fuve{k}}_3 ={ 1 \over q} \oint_{\rm CMBZ}
\frac{\bs dk}{2\pi}\cdot
\bscal{A}\sa\fuve{k} \, .
\end{equation}
Here CMBZ denotes the contour of the MBZ and
the Stoke's theorem was used to write  the second equality.

An expression  similar to eq. (\ref{xyz1}) was originally discussed
by Thouless and Kohmoto \cite{Thou1}.
However, their work is based on the Kubo linear response theory,
consequently only terms linear in  $E$ appear in that case.
Our result reduces to that of Thouless and
Kohmoto if we drop the electric field, i.e.
if we evaluate the conductivity in \Eq{www5}
using the  zero order solution  of  \Eq{pro3}.
As far as the gap condition is satisfied the arguments of  Thouless and Kohmoto can be repeated in order to show that   the expression in 
\Eq{www5} is related to the first Chern number, and consequently  $\eta\sa$ is quantized.
In the case of intermediate electric fiel intensities, altough modified, the  band structure is preserved  \cite{CondMat1} and 
it is  then expected  that the  quantization of the band conductivities should be in principle observable. 
However as the electric field is increased the high density of levels makes it practically impossible 
to resolve the conductance contributions for each miniband, as far as the original Hofstadter miniband 
is concerned this fact can be evaluated as a nonlinear contribution to  $\sigma\sa$ that will lead to a break down 
of its quantization.
Higher order  corrections to the Hall conductivity can be
calculated by a perturbative evaluation of \Eq{pro3}.

We observe that the  term  $-\imath \bs{E}\cdot\nas{k}$
in \Eq{pro3} can be considered as a perturbative potential;
hence a series expansion in $E$ can be worked out. 
The Hall conductance for the $\alpha-$band is expanded as 
\begin{equation}\label{excon}
\sigma\sa =  \sigma\sa^{(0)}  +  \sigma\sa^{(1)}  E  +   \sigma\sa^{(2)}
 E^2 \, + \,  \dots\, ,\,\,\,\,\,\,\,\,\,\,\,\,\,\,\,\,\,\,\,\,\,\,\,\,
 \eta\sa =  \eta\sa^{(0)}  +  \eta\sa^{(1)}  E  +   \eta\sa^{(2)}
 E^2 \, + \,  \dots\, \, .
\end{equation} 

Table~\ref{tabla1} shows the values of $\eta\sa^{(0)}$
 for various  selections of  the inverse magnetic flux $\sigma$.
It is  verified that
$\eta\sa^{(0)}$ always yields an integer value; here $\alpha = (\mu, n)$ labels
the Landau level $\mu$ and the $n$ internal miniband.  For $\sigma=1/3$ we find:
$\eta_{\mu,1}^{(0)}=1$, $\eta_{\mu,2}^{(0)}=-2$, and $\eta_{\mu,3}^{(0)}=1$,
regardless of $\mu$.  The numbers $q$ and $p$ are relatively prime, so there must
be integer numbers $u$ and $v$ such that  $up+vq=1$.  We can identify
$v\equiv \eta \sa^{(0)}$,  hence   $(1-q\eta\sa^{(0)})/p$
is also a integer, but according to  \Eq{cnd4}  this number is  the Hall
conductance of the subband $\alpha$; thus $\sigma\sa^{(0)}$ is quantized in
units of $e^2/h$. 
If the coupling between Landau levels is small, the sum rules
$\sum_n \eta_{\mu,n}^{(0)} =0,$ and $\sum_n \sigma^{(0)}_{\mu,n}=1$
are satisfied guaranteing that the Hall conductivity of  a completely
filled Landau level takes the value $e^2/h$.

As the Fermi energy sweeps through a Landau level, the
partial contributions of the subband (although integer 
multiples of $e^2/h$) do not follow a monotonous behavior.
If the Fermi energy lies in the  $n$-minigap, the accumulated
conductance $\zeta\sa$ is defined as
$\zeta_{\mu,n}=\sum_{j =1}^{j=n}\sigma_{\mu,j}.$ The $n$-minigap conductance
satisfies the Diophantine equation $ n = \lambda \, q + p \, \zeta_{\mu,n} $,
$(\vert \lambda\vert \le p/2)$ in agreement with the results obtained by
Thouless \etal \cite{Thou1}.

\begin{table}[htb]
\caption{Zero and second order coefficients $\eta\sa^{(0)}$, and $\tilde{\eta}\sa^{(2)}$  ($\alpha=(\mu,n)$)
for $\sigma=1/3$, $1/5$ and $1/7$.}\label{tabla1}

\begin{tabular}{c@{$\,\,\,\,$}c@{$\,\,\,\,$}r@{$\,\,\,\,$}|
      @{$\,\,\,\,$}c@{$\,\,\,\,$}c@{$\,\,\,\,$}r@{$\,\,\,\,$}|
      @{$\,\,\,\,$}c@{$\,\,\,\,$}c@{$\,\,\,\,$}r}
\hline
\multicolumn{3}{c}{$\sigma=1/3$, $q=1$, $p=3$}&
\multicolumn{3}{c}{$\sigma=1/5$, $q=1$, $p=5$}&
\multicolumn{3}{c}{$\sigma=1/7$, $q=1$, $p=7$}\\
\hline\hline
 $n$ & ${\eta}^{\left(0\right)}\sa$ & $\tilde{\eta}^{\left(2\right)}\sa$ &
 $n$ & ${\eta}^{\left(0\right)}\sa$ & $\tilde{\eta}^{\left(2\right)}\sa$ &
 $n$ & ${\eta}^{\left(0\right)}\sa$ & $\tilde{\eta}^{\left(2\right)}\sa$\\
\hline
$3$ &  $1$ &  $0.416$ & $5$ &  $1$ &  $-0.018$ & $7$ &  $1$ &    $0.0239$ \\
\hline
    &      &          &     &      &           & $6$ &  $1$ &    $0.2760$ \\
\cline{7-9}
    &      &          & $4$ &  $1$ &  $-4.152$ & $5$ &  $1$ & $-110.2766$ \\
\cline{4-9}
$2$ & $-2$ & $-0.328$ & $3$ & $-4$ &  $17.836$ & $4$ & $-6$ &  $117.3056$ \\
\cline{4-9}
    &      &          & $2$ &  $1$ & $-13.660$ & $3$ &  $1$ &   $-7.3303$ \\
\cline{7-9}
    &      &          &     &      &           & $2$ &  $1$ &    $0.0011$ \\
\hline
$1$ &  $1$ & $-0.087$ & $1$ &  $1$ &  $-0.006$ & $1$ &  $1$ &    $0.0003$ \\
\hline
\end{tabular}
\end{table}
 
We now turn our attention to the higher order correction to the Hall conductance.
Restoring units, the second order contribution to the Hall conductance  has the form 
$\sigma_H^{(2)}  =  \frac{e^3 }{  \hbar V_0^2 \, B}  \, {\tilde \eta}\sa^{(2)}$ where 
\begin{multline}\label{correction22}
{\tilde \eta}\sa^{(2)}  = \sum_{\beta \ne\alpha} \oint_{\rm CMBZ}  d\bs{k} \cdot
\left( \eta\sa^{(0)} -  \eta_\beta^{(0)} \right) h_1\fuve{k}\\
\times
\left[{ \langle \alpha^{(0)} \,
\vert {\partial H \over \partial { k_x} } \vert  \, \beta^{(0)}   \rangle
\langle \beta^{(0)}  \, \vert {\partial H \over \partial {k_i} } \vert \,
\alpha^{(0)}  \rangle \over \left[\cal{E}_{\alpha }-\cal{E}_{\beta}  \right]^4  } \,
+  \,\left(x \leftrightarrow i \right)  \right].
\end{multline}
here the  function ${\bs h}\fuve{k}$ can be selected as the
gauge potential of a unit flux tube in $k-$space:
${\bs h}\fuve{k} = {1 \over 2 \pi} (k_1, - k_2)/(k_1^2 + k_2^2)$.
In Table~\ref{tabla1} we quote some of the values obtained 
for ${\tilde \eta}\sa^{(2)}$.
It is observed that as the internal structure of a  band increases,
the corrections to the conductance of the internal minibands
increases abruptly.
It can be easily proved that the sum rule
$\sum_n \eta_{\mu,n} ^{(2)} =0$ applies.
Consequently, even if each miniband presents
$E^2$  corrections 
to the Hall conductance, the contributions from a completely filled Landau level
cancel exactly,  and the Hall resistance 
quantization remains valid to order ${\cal O} (E^3)$. The nonlinear correction to
$\sigma_H$ are then expected to be dominant for each  miniband and as Table~\ref{tabla1}
suggests these become larger as  the denominator in $\sigma$ increases. 
\section{Summary} 
We address the calculation of the Hall conductance
beyond the linear response approximation. 
A closed expression for the Hall conductance, valid to all orders in $\bs E$ is obtained.
As far as  the  band structure is preserved  the conductivity for each miniband
takes integer values. 
However for intense electric fields the high density of levels makes it practically impossible 
to resolve the conductance contributions for each miniband,  as far as each miniband is concerned
this can be interpretated as nonlinear corrections to $\sigma_H$.
The leading order contribution is quantized  in units of $e^2/h$.
The first order correction  exactly vanishes, while  the  second order
correction shows  a  $\sigma_H^{(2)} \propto e^3 /V_0^2 \, B$
dependence. $\sigma_H^{(2)}$ cancels for a completely filled Landau band.
Hence the nonlinear correction to the conductance is expected to 
be of order  ${\cal O} (E^2)$  for a filled miniband, whereas for a complete
Landau level the correction is expected to be of order
${\cal O} (E^3)$.

\section{Acknowledgements}
We acknowledge the financial support endowed by
CONACyT through grant No. \texttt{42026-F}, the Departamento de
Ciencias B\'asicas UAM-A through project No. \texttt{2230403} and
Rector\'{\i}a General de la UAM-A.

\end{document}